\documentclass[showpacs,prl,onecolumn,aps,superscriptaddress,preprintnumbers,nofootinbib]{revtex4}
\usepackage[T1]{fontenc}
\usepackage[latin9]{inputenc}
\usepackage{amsmath,amssymb}
\usepackage{epsfig}
\usepackage{bm} 
\usepackage{graphicx}
\usepackage{mathrsfs}
\usepackage{amsmath}
\usepackage{amsfonts}
\usepackage{epstopdf}
\usepackage{color}
\usepackage{pifont}
\usepackage{relsize}
\usepackage{mathtools}
\def\slashchar#1{\setbox0=\hbox{$#1$}     		
   \dimen0=\wd0                                 	
   \setbox1=\hbox{/} \dimen1=\wd1               	
   \ifdim\dimen0>\dimen1                        	
      \rlap{\hbox to \dimen0{\hfil/\hfil}}      	
      #1                                        	
   \else                                        	
      \rlap{\hbox to \dimen1{\hfil$#1$\hfil}}   	
      /                                         	
   \fi}

\renewcommand{\vec}{\boldsymbol}
\newcommand{\beq}{\begin{equation}}
\newcommand{\eeq}{\end{equation}}
\newcommand{\bea}{\begin{eqnarray}}
\newcommand{\eea}{\end{eqnarray}}
\newcommand{\ba}{\begin{array}}
\newcommand{\ea}{\end{array}}

\def\eq#1{{Eq.~(\ref{#1})}}
\def\fig#1{{Fig.~\ref{#1}}}
\newcommand{\bas}{\bar{\alpha}_S}
\newcommand{\as}{\alpha_S}
\newcommand{\nn}{\nonumber}

\newcommand{\Lb}{\left(}
\newcommand{\Rb}{\right)}
\newcommand{\h}{\frac{1}{2}}
\newcommand{\rv}{\vec{r}}
\newcommand{\kv}{\vec{k}_T}
\newcommand{\bv}{\vec{b}}
\newcommand{\pom}{I\!\!P}
\newcommand{\xt}{\tilde\xi}
\newcommand{\Y}{\tilde Y}

\begin{document}

\title{ Dipole-dipole scattering amplitude in CGC approach}

\author{Eugene Levin}
\email{leving@tauex.tau.ac.il, eugeny.levin@usm.cl}
\affiliation{Department of Particle Physics, Tel Aviv University, Tel Aviv 69978, Israel}
\affiliation{Departemento de F\'isica, Universidad T\'ecnica Federico Santa Mar\'ia, and Centro Cient\'ifico-\\
Tecnol\'ogico de Valpara\'iso, Avda. Espana 1680, Casilla 110-V, Valpara\'iso, Chile}

\date{\today}


\begin{abstract}
In this paper we propose recurrence relations for the dipole densities in QCD,
which allows us to find these densities from the solution to the BFKL equation. We resolve these relations in the diffusion approximation for the BFKL kernel. Based on this solution, we found the sum of large Pomeron loops. This sum generates  the scattering amplitude that decreases at large values of rapidity $Y$. It turns out that such behaviour of the scattering amplitudes is  an artifact of  diffusion approximation. This approximation leads to the unitarization without saturation both in deep inelastic scattering and in dipole-dipole interaction at high energies.

 \end{abstract}
\maketitle

\vspace{-0.5cm}
\tableofcontents

\section{Introduction}
 The main ideas of Colour Glass Condensate(CGC)/saturation approach (see  Ref.\cite{KOLEB} for a review) : the saturation of the dipole density and the new dimensional scale ($Q_s$), which increases with energy, have become the common language for discussing the high energy scattering in QCD. However, in spite of intensive work 
 \cite{BFKL,LIP,LIREV,LIFT,GLR,GLR1,MUQI,MUDI, MUSA,Salam,NAPE,BART,BKP,MV, KOLE,BRN,BRAUN,BK,KOLU,JIMWLK1,JIMWLK2,JIMWLK3, JIMWLK4,JIMWLK5,JIMWLK6,JIMWLK7,JIMWLK8,AKLL,KOLU1,KOLUD,BA05,SMITH,KLW,KLLL1,KLLL2}, we have several problems, that have not been solved. One of the  principle problems is summing Pomeron loops, without solving which we cannot consider the dilute-dilute and dense-dense parton densities collisions. As has been recently shown\cite{KLLL1,KLLL2}, even the Balitsky-Kovchegov (BK) equation, that governs the dilute-dense parton density  scattering (deep inelastic scattering (DIS) of electron with proton), has to be modified due to  contributions of  Pomeron loops.

  In this paper we  attempt to sum Pomeron loops for dipole-dipole scattering amplitude at high energies. This attempt is based on the experience with the simple, but exactly solvable, two dimensional models\cite{MUSA,ACJ,AAJ,JEN,ABMC,CLR,CIAF,BIT,RS,KLremark2,SHXI,KOLEV,nestor,LEPRI,utm,utmm}, which we will discuss in the next section. From these models we learned,  that the scattering amplitude at high energies is determined by the sum of large Pomeron loops. Actually, the formalism for summing large Pomeron loops in QCD has been developed \cite{MUSA,IAMU,IAMU1,LELU,KO1,LE1}. In this paper, we propose new recurrence relations for the parton densities in QCD, which allows us to find all parton densities from the solution of the BFKL equation. We resolve these recurrence relations in the diffusion approximation for the BFKL kernel and suggest the explicit form for the scattering amplitude. We believe that we have  completed the approach which was started in Refs.\cite{KO1,LE1}.

 
  \section{Pomeron calculus in zero transverse dimension-a recap } 
 The simple toy model: the Pomeron calculus in zero transverse dimension, is a respectable tool  and a well known training ground for the interaction at high energies\cite{ACJ,AAJ,JEN,ABMC,CLR,CIAF,MUSA,Salam,RS,KLremark2,SHXI,KOLEV,BIT,nestor,LEPRI,utm,utmm}. Due to the simplicity of these models, we are able to formulate and solve the reggeon field theory (RFT) for the interacting Pomerons. This theory satisfies both the $s$ and $t$ channel unitarity constraints and includes  the emission of the dipoles as well as  the saturation effects  in the corresponding parton cascades.  The simple toy model also gives  examples of theories that have the probabilistic interpretation for the scattering amplitude in letter and spirit of the partonic approach.  
 
 In Ref.\cite{MUSA} the simple probabilistic formula for the S-matrix is suggested:
  \beq \label{SMS1}
 S(Y) \,\,=\,\,\sum_{n,m} e^{ - m \,n\,\gamma} \,P^{\mbox{\tiny BFKL}}_n( Y_0)\,P^{\mbox{\tiny BFKL}}_m( Y - Y_0)
 \eeq
 where $\gamma $ is the scattering amplitude of two dipoles and $P^{\mbox{\tiny BFKL}}_n( Y) $ is the probability distribution in the BFKL cascade
For $P_n(Y)$  we have equations in the following form for the zero transverse dimension:

\beq \label{PBK}
\frac{ d\, P^{\mbox{\tiny BFKL}}_n(Y)}{ d\,Y} \,\,=\,\, - \Delta\,n\,P^{\mbox{\tiny BFKL}}_n(Y)\,\,+\,\,\Delta\,(n - 1)\,P^{\mbox{\tiny BFKL}}_{n-1}(Y)
\eeq
with the solution:
\beq \label{PNBK}
P_{n}^{BK}(Y)=\frac{1}{N(Y)-1}\left(1-\frac{1}{N(Y)}\right)^{n}\,\,\,\xrightarrow{Y\,\gg\,1}\,\,\,\frac{1}{N(Y)}\exp\Lb - \frac{n}{N(Y)}\Rb
\eeq
where $N(Y)$ is the first factorial  moment or multiplicity of dipoles: $N(Y)= e^{\Delta\,Y}$.
Generally speaking \eq{SMS1} does depend on the reference frame (on the  value of $Y_0$) and, 
as has been discussed in Refs.\cite{MUSA,BIT,utm,utmm}, we need to change \eq{PBK} to obtain the Pomeron calculus which satisfies both $t$ and $s$ channel  unitarity. However, at large values of $Y-Y_0$ and $Y_0$ \eq{SMS1} leads to the scattering amplitude that does not depend on the value of $Y_0$\cite{utmm}.  Indeed,

 \bea \label{SMS2}
\hspace{-0.7cm} S(Y) & =&\sum_{n,m} e^{ - m \,n\,\gamma} \,P^{\mbox{\tiny BFKL}}_n( Y_0)\,P^{\mbox{\tiny BFKL}}_m( Y - Y_0)\,\,=\,\,\sum_{k=0}^\infty 
 \frac{\Lb - \gamma\Rb^k}{k!}\underbrace{\Bigg( \sum_{n=0}^\infty  n^k \,P_n( Y_0) \Bigg)}_{c_k\Lb Y_0\Rb}\underbrace{ \Bigg( \sum_{m=0}^\infty  m^k \,P_m( Y - Y_0) \Bigg)}_{c_k\Lb Y - Y_0\Rb}\,\,\nn\\
 &=& \sum_{k=0}^\infty \,(- \gamma N\Lb Y_0\Rb\,N\Lb Y - Y_0\Rb )^k \,k! \,=\, \sum_{k=0}^\infty \,(- \gamma N\Lb Y\Rb )^k \,k! =\,-\frac{e^{\frac{1}{\gamma  N(Y)}} \text{Ei}\left(-\frac{1}{N(Y) \gamma }\right)}{\gamma  N(Y)} 
\eea
   In \eq{SMS2} we used that (1) $N\Lb Y-Y_0\Rb\,N\Lb Y_0\Rb = N\Lb Y\Rb$; and (2) 
   the factorial moments of the distribution of \eq{PNBK} has the following form
   \beq \label{MS}
   M_k(Y)\,\,=\,\, k!N(Y) \Lb N(Y)\,-\,1\Rb^{k-1} \,\,\xrightarrow{N(Y) \,\gg\,1} \,\,\, c_k(Y) = k! N^k(Y)  
   \eeq
   
   One can see that $S(Y)$, 
   being a function of  $N(Y)$,  does not depend on the reference frame. 
   It turns out that this S-matrix at  large $Y-Y_0$ and $Y_0$  coincides with the one calculated in the UTM\cite{MUSA,utm,utmm}: theory,  which  is  independent  of a reference frame at any value of $Y_0$
 .
 
     \begin{figure}[ht]
    \centering
  \leavevmode
      \includegraphics[width=12cm]{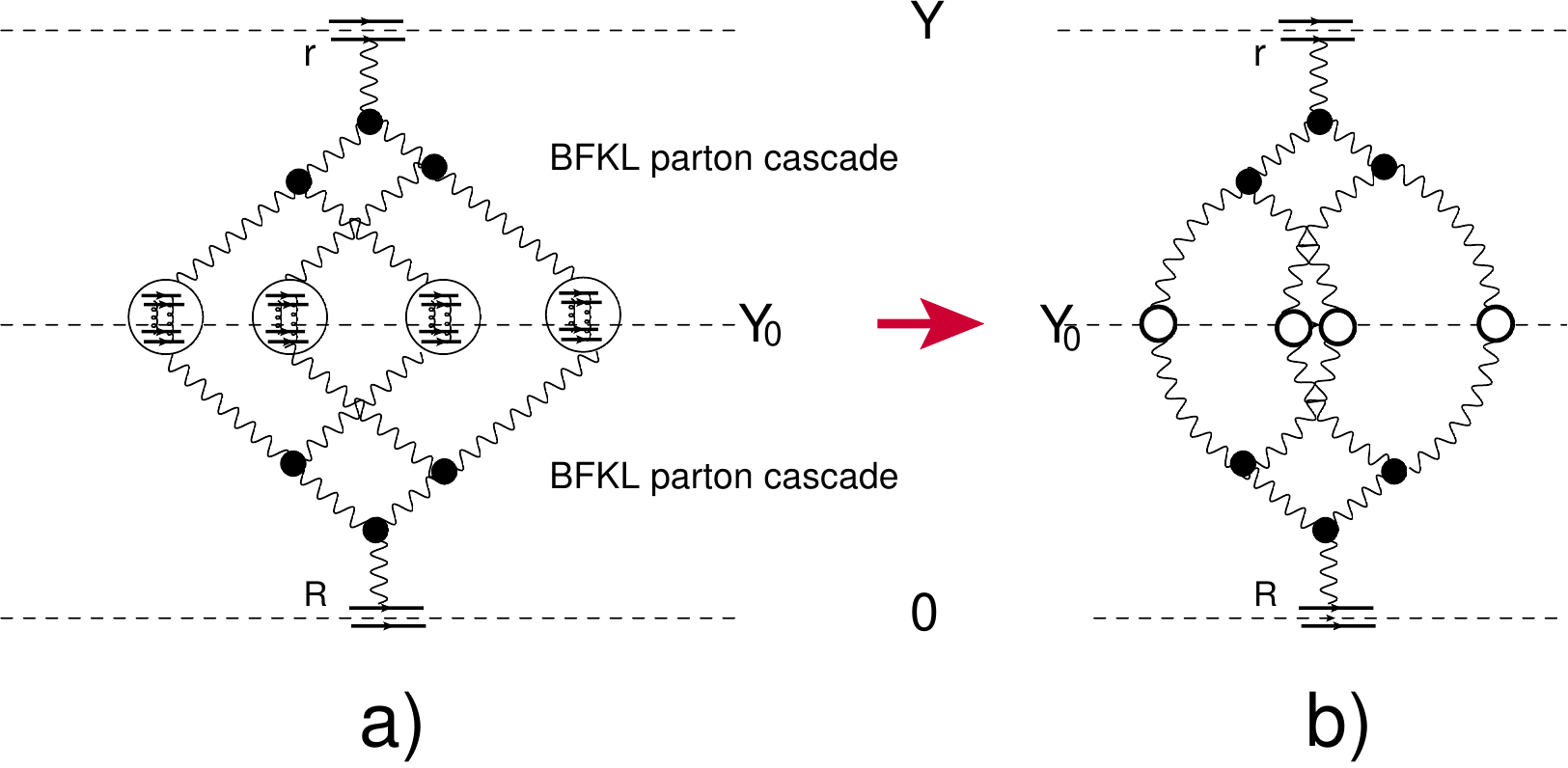}  
      \caption{ Summing  large Pomeron loops. The wavy lines denote the  BFKL Pomeron exchanges in \fig{mpsi}-a and  in \fig{mpsi}-b the Pomeron in the two dimensional Pomeron calculus with
      the Green's function $G_{\pom}(Y) = \exp\Lb \Delta\,Y\Rb$. The black circles stand for the triple Pomeron vertices in both figures, which are equal to $\Delta$ In \fig{mpsi}-b, while the white circles denote the amplitude $\gamma$. In \fig{mpsi}-a we show the dipole-dipole scattering amplitude in the Born approximation of perturbative QCD in the circles.
 }
\label{mpsi}
   \end{figure}
%
It is easy to see that the S-matrix of \eq{SMS1} sums the large BFKL Pomeron loops shown in \fig{mpsi}.  Indeed, the contribution of large Pomeron loops can be written in the following form:\cite{MUDI,IAMU,LELU,KO1,LE1}:
\bea \label{SMS3}
 S(Y) \,\,&=&\,\,\sum_{n}\frac{(- \,\gamma)^n}{n!} M_n\Lb Y - Y_0\Rb\,M_n\Lb Y_0\Rb\,\,\,
 \xrightarrow{ Y- Y_0, Y_0 \gg1} \sum_{n}(- \,\gamma)^n\,n! \Lb N\Lb Y - Y_0\Rb\,N\Lb Y_0\Rb\Rb^n\nn\\
 &=&\,\, \sum_{n}(- \,\gamma)^n\,n! \Lb G_{\pom}\Lb Y - Y_0\Rb\,G_{\pom}\Lb Y_0\Rb\Rb^n  
 = -\frac{e^{\frac{1}{\gamma  N(Y)}} \text{Ei}\left(-\frac{1}{N(Y) \gamma }\right)}{\gamma  N(Y)} 
\eea
 $M_n$ in \eq{SMS3} are the factorial moments which we replace 
 by $M_n (Y)\,=\,n! N^n\Lb Y\Rb$ at large $Y$ for distribution of \eq{PNBK}. In \eq{SMS3} $N\Lb Y\Rb = N\Lb Y- Y_0\Rb\,N\Lb Y_0\Rb$. The advantage of this derivation is  that it can be easily generalize to the QCD case, which is the main goal of this paper.
  
  The factorial moments will play an essential role in our approach. Bearing this in mind, we wish to
  write the equation for them in the simple BFKL cascade of \eq{PBK}. The solution of \eq{PNBK} it is easy to obtain introducing the generating function:
  
  \beq \label{GF}
  Z\Lb Y, u\Rb\,\,=\,\,\sum_{n=1}^\infty\,P_n \Lb Y\Rb \,u^n
  \eeq
  One can see that
  \beq \label{PNMN} 
  P_n\Lb Y\Rb\,\,=\,\,\frac{1}{n!} \frac{\partial\,Z\Lb Y, u\Rb}{\partial u^n}\Big{|}_{u=0} ;~~~
 M_n\Lb Y\Rb\,\,=\,\,\sum_n\Lb  n(n-1)\dots (n-k+1)\Rb\,P_n(Y) = \frac{\partial\,Z\Lb Y, u\Rb}{\partial u^n}\Big{|}_{u=1} ;  
 \eeq

   From \eq{PBK}  the equation for $Z$ takes the form:
   
   \beq \label{ZEQ}
   \frac{\partial\,Z\Lb Y, u\Rb}{ \partial\,Y}\,\,=\,\,- \Delta\,u (1 - u)    \frac{\partial\,Z\Lb Y, u\Rb}{ \partial\,u} 
   \eeq
   Taking $n$-derivatives from \eq{ZEQ} and substituting $u=1$ we obtain the following equation for $M_n(Y)$;
   \beq \label{MNEQ}
   \frac{\partial\,M_n\Lb Y\Rb}{ \partial\,Y}\,\,=\,\,\Delta\,n\,\,M_n\Lb Y\Rb \,\,+\,\Delta\,n (n-1) M_{n-1}\Lb Y\Rb
 \eeq
 \eq{MNEQ} has a more elegant form for $\rho_n(Y)\,\,=\,\,M_n(Y)/n! $:
  \beq \label{RHONEQ}
   \frac{\partial\,\rho_n\Lb Y\Rb}{ \partial\,Y}\,\,=\,\,\Delta\,n\,\,\rho_n\Lb Y\Rb \,\,+\,\Delta\,(n-1) \rho_{n-1}\Lb Y\Rb
 \eeq  
  For sum of the large Pomeron loops we have the following formula, using $\rho_n(Y)$:
  \beq \label{SMS4}
 S(Y) \,\,=\,\,\sum_{n}(- \,\gamma)^n\,n! \, \rho_n\Lb Y - Y_0\Rb\,\rho_n\Lb Y_0\Rb
\eeq   
   \eq{SMS4} has been generalized to the QCD case in Refs.\cite{IAMU,LELU}    and we will use it in our approach. 
     
     \eq{ZEQ} for the generating function $Z$ can be rewritten as the non-linear equation for $Z$ in the form:
     \beq \label{ZNEQ}
     \frac{\partial\,Z\Lb Y, u\Rb}{ \partial\,\Delta\,Y}\,\,=\,\,-Z\Lb Y, u\Rb \Big( 1\,\,-\,\,Z\Lb Y, u\Rb\Big)
     \eeq    
     Bearing in mind that $\rho_n\Lb Y\Rb\,\,=\,\,\frac{1}{n!} \frac{\partial\,Z\Lb Y, u\Rb}{\partial u^n}\Big{|}_{u=1}$ we can obtain the equation for $\rho_n$ differentiating \eq{ZNEQ} and puting $u=1$ :
      \beq \label{RHONN}
     \frac{\partial\,\rho_n\Lb Y \Rb}{ \partial\,\Delta\,Y}\,\,=\,\,\rho_n\Lb Y \Rb\,\,+\,\,\sum_{k=1}^{n-1}\rho_{n-k}\Lb Y \Rb\,\rho_k\Lb Y\Rb
     \eeq 
     Subtracting this equation from \eq{RHONEQ} we obtain the following  recurrence relation for $\rho_n$:
     \beq \label{RECR}
    \Lb n\,-\,1\Rb\, \rho_n\Lb Y \Rb\,\,=\,\,\sum_{k=1}^{n-1}\,  \rho_{n - k}\Lb Y \Rb\,\rho_{k}\Lb Y\Rb \,\,-\,\,\Lb n-1\Rb\,\rho_{n-1}\Lb Y\Rb     \eeq      
     Solution to \eq{RECR} has the form:
     
     \beq \label{RHONREC}
     \rho_n\Lb Y\Rb\,\,=\,\,\rho_1\Lb Y \Rb \Big( \rho_1\Lb Y \Rb\,\,-\,\,1\Big)^{n - 1}
     \eeq    
     
     Summarizing what we have obtained in this section for the simple models of RTF, we conclude that (i)  the scattering amplitude at  high energies can be calculated from \eq{SMS4} using the parton densities $\rho_n(Y)$; (ii) these parton densities satisfy the two evolution equations of \eq{RHONEQ} and \eq{RHONN};  and (iii) these two equations lead to the recurrence  relation for $\rho_n$  (see \eq{RECR}. The main goal of this paper is to generalize these ingredients to the case of QCD and obtain the QCD scattering amplitude at high energies.    
      
    
    \section{Summing large Pomeron loops in QCD} 

As it has been mentioned, our main goal is to sum large Pomeron loops in QCD to obtain the scattering amplitude. Our approach  includes two stages. First, we need to generalize   \eq{SMS4} to the case of QCD. Actually, this problem has been solved in Refs.\cite{MUSA,Salam,IAMU,KOLEB,MUDI,LELU} and we are going to discuss it here.
Second, we need to get the evolution equations for the parton densities and find their solutions. This problem has been partly solved in Ref.\cite{LELU}, but in this section we will find the second evolution equation  and suggest the recurrence relations for $\rho_n$. Finally, we need to find the solution for the parton densities and this topic is the main one of  this section.


     \subsection{BFKL Parton cascade:  evalution  equations  and recurrence relations  for the parton densities}

     The simple \eq{PBK} has been generalized to the QCD case in  
      Refs.\cite{KOLEB,MUDI,LELU}   and has  the following form :
  \beq  \label{PC1}
\frac{\partial\,P_n\Lb Y, \vec{r }, \vec{b};\{\vec{r}_i,\vec{ b}_i\} \Rb}{ 
\partial\, Y }\,=\,-\,
\sum^n_{i=1}\,\omega_G(r_i) \,
P_n\Lb Y, \vec{r }, \vec{b};\,\{\vec{r}_i,\vec{ b}_i\},\Rb \,\,+\,\,\bas\,\sum^{n-1}_{i=1} \,\frac{(\vec{r}_i\,+\, 
\vec{r}_n)^2}{(2\,\pi)\,r^2_i\,r^2_n}\,
P_{n - 1}\Lb Y, \vec{r},\vec{b};\{\vec{r}_j, \vec{b}_j ,\vec{r}_i+ \vec{r}_n,\vec{b}_{in}\}
\Rb\nn
\eeq
  where $P_n\Lb Y, r, b ; \{r_i,b_i\}\Rb$ is the probability to have $n$-dipoles
 of size $r_i$,  at impact parameter $b_i$ and  at rapidity $Y$
  . $\vec{b}_{in} $ in \eq{PC1} is equal to $\vec{b}_{in}
 \,=\,\vec{b}_i \,+\,\h \vec{r}_i \,=\,\vec{b}_n \,-\,\h \vec{r}_i$.
  
  \eq{PC1} is a typical cascade equation in which the first term
 describes the reduction   of  the probability to find $n$ dipoles
 due to the possibility that one of $n$ dipoles can decay into two dipoles 
of
 arbitrary sizes  
  , while the second term,  describes  the growth due to the 
splitting
 of $(n-1)$ dipoles into $n$ dipoles.   
  We introduce the generating functional\cite{MUDI}

\beq \label{Z}
Z\Lb Y, \vec{r},\vec{b}; [u_i]\Rb\,\,=\,\,\sum^{\infty}_{n=1}\int P_n\Lb Y,\vec{r},\vec{b};\{\vec{r}_i\,\vec{b}_i\}\Rb \prod^{n}_{i=1} u\Lb \vec{r}_i\,\vec{b}_i\Rb\,d^2 r_i\,d^2 b_i
\eeq
 where $u\Lb \vec{r}_i\,\vec{b}_i\Rb \equiv\,u_i$ is an arbitrary function.
 The initial and  boundary conditions for \eq{Z}  take 
the following form for the functional $Z$:
\begin{subequations}
\bea
Z\Lb Y=0, \vec{r},\vec{b}; [u_i]\Rb &\,\,=\,\,&u\Lb \vec{r},\vec{b}\Rb;\label{ZIC}\\
Z\Lb Y, r,[u_i=1]\Rb &=& 1; \label{ZSR}
\eea
\end{subequations}

Multiplying both parts of \eq{PC1} by $\prod^{n}_{i=1} u\Lb \vec{r}_i\,\vec{b}_i\Rb$ and integrating over $r_i$ and $b_i$ we obtain the following linear functional equation\cite{LELU};
\begin{subequations}
\bea
&&\hspace{-0.7cm}\frac{\partial Z\Lb Y, \vec{r},\vec{b}; [u_i]\Rb}{\partial \,Y} =\int d^2 r'\,  K\Lb \vec{r}',\vec{r} - \vec{r'}|\vec{r}\Rb\Bigg( - u\Lb r, b\Rb\,\,+\,\,u\Lb \vec{r}',\vec{b} + \h(\vec{r} - \vec{r}') \Rb \,u\Lb \vec{r} - \vec{r}',\vec{b} - \h\vec{r}'\Rb\Bigg) \frac{\delta\,Z}{\delta \,u\Lb r, b \Rb};\label{EQZ}\\
&&  K\Lb \vec{r}',\vec{r} - \vec{r'}|\vec{r}\Rb\,=\frac{\bas}{2 \,\pi}\frac{r^2}{r'^2\,(\vec{r} - \vec{r}')^2} ;\,~~~~~
 \omega_G\Lb r\Rb\,\,=\,\,\int d^2 r'  K\Lb \vec{r}',\vec{r} - \vec{r'}|\vec{r}\Rb; \label{OMG}
 \eea
\end{subequations}
     The $n$-dipole densities 
$\rho_n(r_1, b_1,\ldots\,,r_n, b_n)$
 are defined as follows\cite{LELU}:
\beq \label{N2}
\rho_n(r_1, b_1\,
\ldots\,,r_n, b_n; Y\,-\,Y_0)\,=\,\frac{1}{n!}\,\prod^n_{i =1}
\,\frac{\delta}{\delta
u_i } \,Z\left(Y\,-\,Y_0;\,[u] \right)|_{u=1}
\eeq
Taking n-th functional derivatives from \eq{EQZ}  and substituting $u_1=1$ we  obtain for 
 $\rho_n$ \cite{LELU} :
\bea \label{N3}
\frac{\partial \,\rho_n(\{r_i, b_i\})}{ 
\,\partial\,Y}\,\,&=&\,\
-\,\sum_{i=1}^n
 \,\,\omega_G(r_i)\,\,\rho_n(\{r_i, b_i\})\,\,+\,\,2\,\sum_{i=1}^n\,
\int\,\frac{d^2\,r'}{2\,\pi}\,
\frac{r'^2}{r^2_i\,(\vec{r}_i\,-\,\vec{r}')^2}\,
\rho_n(\ldots\,r', b_i-r'/2\dots)\nn\\
 & & 
\,
+\bas \frac{(\vec{r}_i\,+\,\vec{r}_n)^2}{ r^2_i\,\,r^2_n}\,\sum_{i=1}^{n-1}
\rho_{n-1}(\ldots\,(\vec{r}_i\,+\,\vec{r}_n), b_{in}\dots).
\eea

Introducing 
\beq \label{BRHO}
\bar{\rho}_n(\rv, \bv; \{r_i, b_i\}) \,=\,\,\prod_{i=1}^n \,r^2_i\,\,\rho_n(\{r_i, b_i\})
\eeq
we reduce \eq{N3} to the following form
\bea \label{N30}
\frac{\partial \,\bar{\rho}_n(\{\rv_i, \bv_i\})}{ 
\,\partial\,Y}\,\,&=&\,\,\sum_{i=1}^n\,
\int\,\frac{d^2\,r'}{2\,\pi}\,
K\Lb \vec{r}',\vec{r}_i - \vec{r'}|\vec{r}_i\Rb\\
&\times&\Bigg\{ \bar\rho_n(\{\rv_j,\bv_j\}, \rv',\bv_i - (\rv_i - \rv')/2)
\,+\,\bar\rho_n(\{\rv_j,\bv_j\}, \rv_i -  \rv',\bv_i - \rv'/2)\,-\,\bar{\rho}_n(\{\rv_i, \bv_i\})\Bigg\}
\nn\\
 & & 
\,
+\,\bas\sum_{i=1}^{n-1}\,
\bar\rho_{n-1}(\ldots\,(\vec{r}_i\,+\,\vec{r}_n), b_{in}\dots).\nn
\eea

\eq{EQZ} can be rewritten as the non-linear equation for $Z$\cite{MUDI,LELU}:

\beq
\hspace{-0.7cm}\frac{\partial Z\Lb Y, \vec{r},\vec{b}; [u_i]\Rb}{\partial \,Y} =\int d^2 r'\,  K\Lb \vec{r}',\vec{r} - \vec{r'}|\vec{r}\Rb\Bigg( - Z\Lb Y, \vec{r},\vec{b}; [u_i]\Rb   \,\,+\,\,Z\Lb \vec{r}',\vec{b} + \h(\vec{r} - \vec{r}'); [u_i]\Rb \,Z\Lb \vec{r} - \vec{r}',\vec{b} - \h\vec{r}'; [u_i]\Rb\Bigg) \label{EQNZ}\\
\eeq
 Using the definition of \eq{N2} and differentiating  \eq{EQNZ}  (see \eq{N2})we obtain a new equation for $\rho_n$\footnote{The analogous equation for the factorial moments of multiplicity distribution has been derived in Refs.\cite{LMM, LEM}.} :
 \bea \label{RHONNEQ}
 &&\frac{\partial \,\rho_n(\rv, \bv; \{\rv_i, \bv_i\})}{ 
\,\partial\,Y}\,\,=\,\, \bas\,\int d^2 r'\,  K\Lb \vec{r}',\vec{r} - \vec{r'}|\vec{r}\Rb\\
& \times&\,\,\Bigg\{\Lb\rho_n(\rv', \bv- \h(\rv - \rv'); \{\rv_i, \bv_i\}) \,\,+\,\,\rho_n(\rv - \rv', \bv- \h\rv'; \{\rv_i, \bv_i\}) -\,\,\rho_n(\rv, \bv; \{\rv_i, \bv_i\})\Rb\nn\\
 &+&\bas\,\,\sum^{n-1}_{k=1} \rho_{n-k}(\rv', \vec{b} + \h(\vec{r} - \vec{r}'; \{\rv_i, \bv_i\}) \,\rho_k(\rv -\rv', \bv - \h \rv'; \{\rv_i, \bv_i \})\Bigg\} \nn
 \eea

 For $\bar \rho_n$ it has the same form.
This equation together with \eq{N30} leads to the recurrence relation for $\bar \rho_n$, which has the form:

 \bea \label{N301}
&&\int d^2 r'\,  K\Lb \vec{r}',\vec{r} - \vec{r'}|\vec{r}\Rb\,\Bigg\{\Lb\rho_n(\rv', \bv- \h(\rv - \rv'); \{\rv_i, \bv_i\}) \,\,+\,\,\rho_n(\rv - \rv', \bv- \h\rv'; \{\rv_i, \bv_i\}) -\,\,\rho_n(\rv, \bv; \{\rv_i, \bv_i\})\Rb\nn\\
 &+&\,\sum^{n-1}_{k=1} \rho_{n-k}(\rv', \vec{b} + \h(\vec{r} - \vec{r}'; \{\rv_i, \bv_i\}) \,\rho_k(\rv -\rv', \bv - \h \rv'; \{\rv_i, \bv_i\}) \Bigg\}\,\,=\,\,
  \sum_{i=1}^n\,
\int\,\frac{d^2\,r'}{2\,\pi}\,
K\Lb \vec{r}',\vec{r}_i - \vec{r'}|\vec{r}_i\Rb \nn\\
&\times&\Bigg\{ \bar\rho_n(\rv, \{\rv_j,\bv_j\}, \rv',\bv_i - (\rv_i - \rv')/2)
\,+\,\bar\rho_n(\rv,\{\rv_j,\bv_j\}, \rv_i -  \rv',\bv_i - \rv'/2)\,-\,\bar{\rho}_n(\rv,\{\rv_i, \bv_i\})\Bigg\}
\nn\\
 &+ & 
\,
\,\,\sum_{i=1}^{n-1}\,
\bar\rho_{n-1}(\rv, \ldots\,(\vec{r}_i\,+\,\vec{r}_n), b_{in}\dots)
\eea

 One can see that just from general form of \eq{N30} the leading energy behaviour stems from the  inhomogeneous term of this equation and we expect that $\bar \rho_n\,\propto\,\sum^{n-1} _{k} \,\bar \rho_{n-k}\,\bar \rho_k$.
 
     
     \subsection{Main formula}
      The scattering amplitude shown in \fig{mpsi}-a can be written in the following way \cite{LELU,KO1,LE1} :
      
      \beq \label{MPSI}
      A\Lb Y, r,R ;  \vec{b}\Rb\,=\,\sum^\infty_{n=1}\,\Lb -1\Rb^{n+1}\,n!\int  \prod \frac{d^2 r_i}{r^4_i}\,\frac{d^2\,r'_i}{r'^4_i }\,d^2 b'_i 
     \int d^2 \delta b_i \gamma^{BA}\Lb r_1,r'_i, \vec{b}_i -  \vec{b'_i}\equiv \delta \vec{b} _i\Rb 
    \,\,\bar{\rho}\Lb Y - Y_0, \{ \vec{r}_i,\vec{b}_i\}\Rb\,\bar{\rho}\Lb Y_0, \{ \vec{r}'_i,\vec{b}'_i\}\Rb   \eeq
  $\gamma^{BA}$ is the scattering amplitude of two dipoles in the Born approximation of perturbative QCD. Considering $Y - Y_0 \gg 1$ and $Y_0 \gg 1$   one can see that      the typical $\ln b^2_i \,\sim\, \sqrt{Y - Y_0}$ and $\ln b'^2_i \,\sim\, \sqrt{Y_0}  $ are large, while in  $\gamma^{BA}$ $\delta b_i \,\sim\,r_1, r'_i$. Hence, we can neglect the contribution of  $ \delta b_i$ in $\bar\rho$. Therefore, in \eq{MPSI} enters 
    $  \int d^2 \delta b_i \gamma^{BA}\Lb r_1,r'_i, \vec{b}_i -  \vec{b'_i}\equiv \delta \vec{b} _i\Rb $ which can be written as\cite{MUDI}:
    \beq \label{BAAMPL}
 \sigma^{\rm BA}\Lb \rv_i, \rv'_i\Rb\,\,=\,\,    \int d^2 \delta b_i \gamma^{BA}\Lb r_1,r'_i, \vec{b}_i -  \vec{b'_i}\equiv \delta \vec{b} _i\Rb\,\,=\,\,4\,\pi\,\bas^2\int \frac{d\,l}{l^3}\,\Lb 1\,-\,J_0\Lb l\,r_i\Rb\Rb\,\,  \Lb 1\,-\,J_0\Lb l\,r'_i\Rb\Rb  
  \eeq  
   In \eq{MPSI} $\vec{b}_i\,\,=\,\,\vec{b} \,-\,\vec{b'}_i$. 
     \begin{boldmath}
     \subsection{Solutions for $\bar{\rho}_n\Lb Y, \{r_i,b_i\}\Rb$}
      \end{boldmath}     
     \begin{boldmath}
     \subsubsection{ $\bar{\rho}_1\Lb Y, r, r_1,b_1\Rb$}
      \end{boldmath}     

From 
\eq{N30} for $\bar\rho_1$ 
we have  the linear equation: 
\beq \label{N4}
\frac{\partial \,\bar\rho_1(Y; r_1, b)}{ 
\bas\,\partial\,Y}\,\,=\,\,-\,\,\omega_G\Lb r_1\Rb\bar\rho_1(Y; r_1, b) \,\,+\,\,2\,\int\,\frac{d^2\,r'}{2\,\pi}\,
K\Lb \vec{r}',\vec{r} - \vec{r'}|\vec{r}\Rb\,
\bar{\rho}_1\Lb Y, r',b\Rb
\eeq
The physical meaning of $\rho_1$ is clear from \eq{N2}: it is the mean
 number of dipoles with size $r_1$  in the partonic wave function of the 
 projectile or target.    
      It is proven in Ref.\cite{LIP} that the eigenfunction of the BFKL equation has the following form

\beq \label{EIGENF}
\phi_\gamma\Lb \vec{r} , \vec{r}_1, \vec{b}_1\Rb\,\,\,=\,\,\,\Lb \frac{ r^2\,r_1^2}{\Lb \vec{b}_1  + \h(\vec{r} - \vec{r}_1)\Rb^2\,\Lb \vec{b}_1  -  \h(\vec{r} - \vec{r}_1)\Rb^2}\Rb^\gamma\,\,\xrightarrow{b_1\,\gg\,r,r_1}\,\,\Lb \frac{ r^2\,r_1^2}{b_1^4}\Rb^\gamma\,\,\equiv\,\,e^{\gamma\,\xi}\eeq
for any kernel which satisfies the conformal symmetry. In \eq{EIGENF} $r$ is the size of the initial dipole at $Y=0$  while $r_1$ is the size of the dipole with rapidity $Y$. As has been discussed in the previous section, the typical $b_i$ in $\bar{\rho}_n$ in \eq{MPSI} is large. Hence we can use the variable $\xi$ from \eq{EIGENF}.

For the kernel of the LO BFKL equation (see \eq{OMG}) the eigenvalues  take the form:

\beq \label{CHI}
\omega\Lb \bas, \gamma\Rb\,\,=\,\,\bas\,\chi\Lb \gamma \Rb\,\,\,=\,\,\,\bas \Lb 2 \psi\Lb 1\Rb \,-\,\psi\Lb \gamma\Rb\,-\,\psi\Lb 1 - \gamma\Rb\Rb\,\,\xrightarrow{\gamma\,\to\,\h}
\,\underbrace{4\ln2 \bas}_{\Delta_{\mbox{\tiny{BFKL}}}}\,\,+\,\,\underbrace{14 \zeta(3) \bas}_{D}\Lb \gamma - \h\Rb^2
\eeq
where $\psi(z)$  is the Euler psi-function $\psi\Lb z\Rb = d \ln \Gamma(z)/d z$,    $\bas = N_c \as/\pi$, where $N_c$ is the number of colours. The general solution to \eq{N4} takes the form:

\beq \label{SOLN1}
\bar\rho_1(Y; r_1, b)\,\,=\,\,\int\limits^{\epsilon + i \infty}_{\epsilon - i \infty} \frac{d\,\gamma}{2\,\pi\,i} 
\int\limits^{\epsilon + i \infty}_{\epsilon - i \infty} \frac{d\,\omega}{2\,\pi\,i}\frac{1}{\omega\,-\, \omega\Lb \bas, \gamma\Rb} e^{\omega\,Y\,\,+\,\,\gamma\,\xi} \phi_{in}\Lb \gamma\Rb\,\,=\,\,
  \,\int\limits^{\epsilon + i \infty}_{\epsilon - i \infty} \frac{d\,\gamma}{2\,\pi\,i} 
 e^{\omega\Lb \bas, \gamma\Rb\,Y\,\,+\,\,\gamma\,\xi} \phi_{in}\Lb \gamma\Rb   
   \eeq  
    Function $\phi_{in}\Lb \gamma\Rb$ has  been found from the initial conditions at $Y=0$\cite{LIP}(see also Refs.\cite{MUDI,LIREV}): $\phi_{in} \,=\,i \nu/\pi$, where $\gamma = \h + i \nu$.   For large $Y$ we can estimate the integral over $\gamma$ using the steepest descent method. The equation for the saddle  point has the form
    
    \beq \label{SP} 
      \frac{d \,\omega\Lb \bas, \gamma\Rb}{d\,\gamma } \,Y|_{\gamma=\gamma_{\rm SP}}\,\,+\,\,\xi\,\,= 0
      \eeq
 The solution to \eq{SP} gives $\gamma_{SP} \,=\,\h\,\,- \,\frac{\xi}{2\,D\,Y}\,\xrightarrow{Y\,\gg\,\xi}\,\h\, $.        
         
         Plugging \eq{SP} in \eq{SOLN1} and using $\phi_{in} \,=\,i \nu_{\rm SP}/\pi$,  we obtain:
   \beq \label{SOLN1F}
\bar\rho^{\rm d.a.}_1(Y; r,  r_1, b)\,\,=\,\,\frac{2}{\Lb D\,Y\Rb^{3/2}}\,\xi\,e^{\Delta_{\mbox{\tiny{BFKL}}}\,Y-\frac{\xi^2}{4\,D\,Y}}\,e^{\h \xi}\,\,=\,\,\,\frac{2}{\Lb D\,Y\Rb^{3/2}}\frac{r\,r_1}{b^2_1}\,\ln\Lb \frac{b^4_1}{r^2\,r^2_1} \Rb\,\exp\Lb  \Delta_{\mbox{\tiny{BFKL}}}\,Y-\frac{ \ln^2\Lb\frac{r^2\,r^2_1}{b^4_1}\Rb}{ 4\,D\,Y}\Rb
\eeq        
    where d.a. denotes diffusion approximation for the BFKL kernel in the vicinity of $\gamma=\h$(see \eq{OMG}), which has been used in deriving \eq{SOLN1F}. It is instructive to note that for $\bar\rho_1(Y; r_1, b)$    \eq{RHONNEQ} has the same form as \eq{N30} and the same solution as \eq{SOLN1F}.
    We will use below $\bar{\rho}_n$ in the momentum representation: viz.
     \beq \label{MMOM}
\bar \rho_n\Lb Y,\vec{k}_T, \vec{b}; Y, \{\rv_i,\bv_i\}\Rb \,\,=\,\,\int d^2 r\,e^{- i \vec{k}_T \cdot \,\vec{r}} \frac{\bar\rho_{n}(Y,\rv, \vec{b} ; \{\rv_i,\bv_i\})}{r^2}.
\eeq   

It turns out that in the vicinity of $\gamma_i=\h$  we can obtain the momentum representation of $\bar \rho_n$ (and vise versa) 
using simple substitute: $ r_i\,\, \rightleftarrows \,\,k^i_T/2$ (see Ref.\cite{RY} formula {\bf 6.561(14)}). Hence,
    \beq \label{SOLN1FM}
G_{\pom}(Y; k_T,  r_1, b)\,\equiv\,\bar\rho^{\rm d.a.}_1(Y; k_T,  r_1, b)\,\,=\,\,\frac{2}{\Lb D\,Y\Rb^{3/2}}\,\xi'\,e^{\Delta_{\mbox{\tiny{BFKL}}}\,Y-\frac{\xi'^2}{4\,D\,Y}}\,e^{\h \xi'}
\eeq   
  with $\xi' = \ln \Lb\frac{4\,r^2_1}{k^2_T\,b^4_1}\Rb$. In conclusion, we see that  $\bar \rho_1$ is described by the exchange of the BFKL Pomeron, which in diffusion approximation has the form of \eq{SOLN1FM}.

     \begin{boldmath}
     \subsubsection{ $\bar{\rho}_2\Lb Y, r, r_1,b_1, r_2,b_2\Rb$}
      \end{boldmath}     
    
    For $\bar{\rho}_2\Lb Y, \rv; \rv_1,b_1, \rv_2,b_2\Rb$      \eq{N301} can be rewritten as follows:
   \bea \label{N20}
&&\int\,\frac{d^2\,r'}{2\,\pi}\,
K\Lb \vec{r}',\vec{r}_1 - \vec{r'}|\vec{r}_1\Rb\,
\Big\{\bar{\rho}_2\Lb Y,\rv;  \vec{r}',b_1,\vec{r}_2, b_2\Rb\,\,+\,\,\bar{\rho}_2\Lb Y,\rv; \vec{r_1} - \vec{r}',b_1,\rv_2, b_2\Rb\,-\, \bar\rho_2(Y, \rv; \vec{r}_1, b_1,\vec{r}_2, b_2)\Big\}\nn\\
&&
\,\,+\,\int\,\frac{d^2\,r'}{2\,\pi}\,
K\Lb \vec{r}',\vec{r}_2 - \vec{r'}|\vec{r}_2\Rb\,
\Big\{\bar{\rho}_2\Lb Y,\rv;  \vec{r}_1,b_1,\vec{r}', b_2\Rb\,\,+\,\,\bar{\rho}_2\Lb Y,\rv; \vec{r}_1,b_1,  \vec{r}_2 - \vec{r}',b_2\Rb\,-\, \bar\rho_2(Y, \rv; \vec{r}_1, b_1,\vec{r}_2, b_2)\Big\}\nn\\
&+&\rho_1(Y, \rv;  \vec{r}_1 +\vec{r}_2, b)\nn\\
&=&\,\,\int \frac{d^2\,r' }{ 2\,\pi} K\Lb \vec{r}',\vec{r} - \vec{r'}|\vec{r}\Rb\nn\\
 &\times&\Bigg\{ \Bigg(\,\bar \rho_2(Y,\rv', \bv -\h(\rv - \rv'); \rv_1, b_1, \rv_2, b_2)\,\,+\,\,
 \bar \rho_2(Y,\rv - \rv', \bv -\h \rv'; \rv_1, b_1, \rv_2, b_2) \,\,-\,\, \bar \rho_2(Y,\rv , \bv ; \rv_1, b_1, \rv_2, b_2) \Bigg)\nn\\
 &+&\,\,\, \bar\rho_{1}(Y,\rv', \vec{b} ; \rv_1, \bv_1) \,\bar\rho_1(Y,\rv -\rv', \bv; \rv_2, \bv_2) \Bigg\}
 \eea

    In \eq{N2}  we neglect the shifts in the impact parameters due to the sizes of dipoles since in \eq{MPSI} all $b_i$ are much larger then $r_i$.

The simplest solution to \eq{N2} we obtain in diffusion approximation for the BFKL kernel (see \eq{CHI}). In this approximation  
\bea \label{DAEQ}
&&\int\,\frac{d^2\,r'}{2\,\pi}\,
K\Lb \vec{r}',\vec{r}_1 - \vec{r'}|\vec{r}_1\Rb\,
\Big\{\bar{\rho}_2\Lb Y,\rv;  \vec{r}',b_1,\vec{r}_2, b_2\Rb\,\,\,+\,\,\bar{\rho}_2\Lb Y,\rv; \vec{r}_1,b_1,  \vec{r}_2 - \vec{r}',b_2\Rb\,-\, \bar\rho_2(Y, \rv; \vec{r}_1, b_1,\vec{r}_2, b_2)\Big\}\nn\\
&=&\,\Bigg( \Delta_{\mbox{\tiny BFKL}}\,\,+\,\,D\,\frac{\partial^2}{\partial  \,\xi_1^2}\Bigg) \bar{\rho}_2\Lb Y,\rv;  \rv_1,b_1,\vec{r}_2, b_2\Rb\,\equiv\,L^{d.a.} \bar{\rho}_2\Lb Y,\rv;  \rv_1,b_1,\vec{r}_2, b_2\Rb\nn
\eea
for all $r_i$ and $r$ in \eq{N2}.

\begin{figure}[ht]
    \centering
  \leavevmode
      \includegraphics[width=9cm]{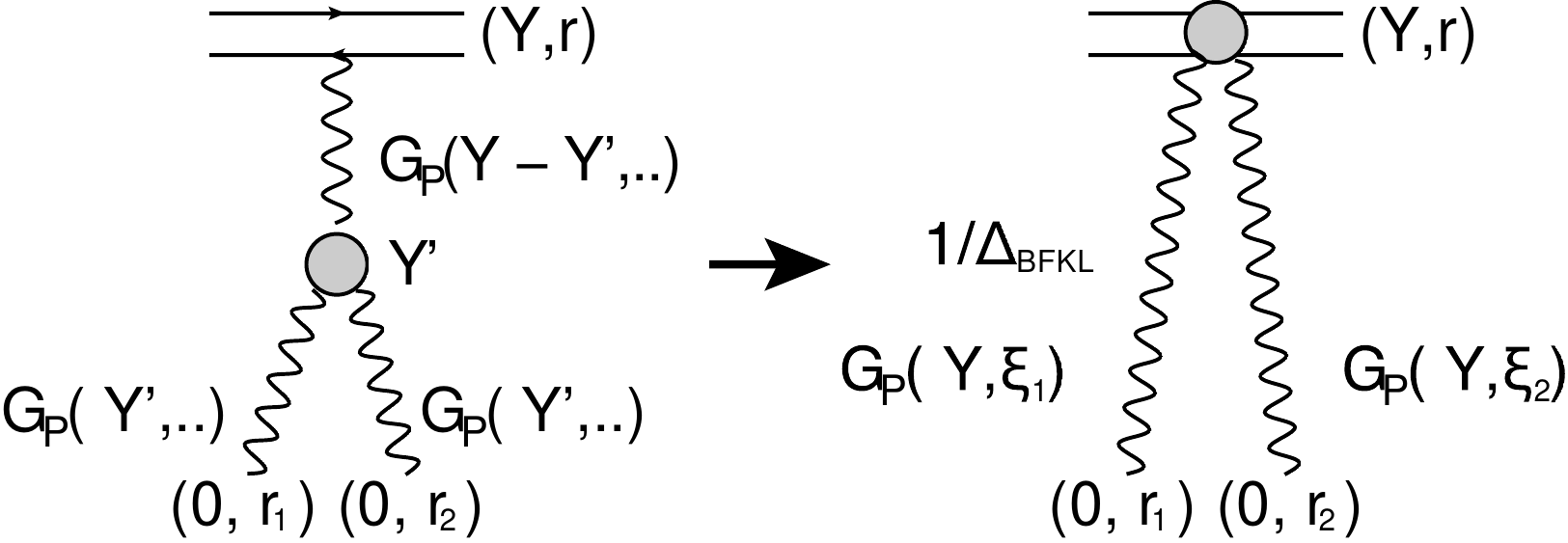}  
      \caption{ The graphic form of \eq{RHO2F}. The wavy lines describe the BFKL Pomerons. The blob corresponds to triple Pomeron vertex. Factor $1/\Delta_{\mbox{\tiny BFKL}} $ stems from integration over $Y'$ the triple Pomeron diagram.}
\label{pom2}
   \end{figure}

We resolve the recurrence relation of \eq{N2} by
 neglecting all contributions of the order of $ \frac{\xi_i^2}{4\,D\,Y^2}$   in kernels $K$, replacing them by $\Delta_{\mbox{\tiny BFKL}}$ (see \eq{SP}).  Indeed, in this case 

\bea \label{RHON2F1}
&&\bar\rho_2(Y, \rv; \vec{r}_1, b_1,\vec{r}_2, b_2)\,\,=\\
&&\,\,\frac{\bas}{\Delta_{\mbox{\tiny BFKL}}}\Bigg\{ \int d^2\,r'  K\Lb \vec{r}',\vec{r} - \vec{r'}|\vec{r}\Rb\bar\rho_{1}(Y, \rv', \vec{b} ; \rv_1, \bv_1) 
\rho_{1}(Y,\rv - \rv', \vec{b} ; \rv_2, \bv_2)\,\,-\,\,\bar\rho_1(Y, \rv;  \vec{r}_1 +\vec{r}_2, b\Bigg\}\nn\eea

\eq{RHON2F1} can be rewritten in more economic form going to the momentum representation (see \eq{MMOM}):
\beq \label{RHON2F2}
\bar\rho_2(Y, \kv; \vec{r}_1, b_1,\vec{r}_2, b_2)\,\,=\,\,
\,\,\frac{\bas}{\Delta_{\mbox{\tiny BFKL}}}\Bigg\{ \bar\rho_{1}(Y, \kv, \vec{b} ; \rv_1, \bv_1) 
\rho_{1}(Y,\kv, \vec{b} ; \rv_2, \bv_2)\,\,-\,\,\bar\rho_1(Y, \kv;  \vec{r}_1 +\vec{r}_2, b\Bigg\}\eeq

\eq{RHON2F2}  leads to the following estimates in the diffusion approximation:
  
  \bea \label{RHO2F}
\bar\rho_2(Y, \xi'_1,\xi'_2)& =&\Lb \frac{\bas}{\Delta_{\mbox{\tiny BFKL}}}\Rb \Bigg\{\bar\rho^{\rm d.a.}_1\Lb Y;  \xi'_1\Rb \bar\rho^{\rm d.a.} _1\Lb Y;  \xi'_2\Rb \,\,-\,\, \bar\rho^{\rm d.a.}_1\Lb Y;  \xi'_{12}\Rb\Bigg\} \\
&\,\,\xrightarrow{Y \gg 1} &\,\,\Lb \frac{\bas}{\Delta_{\mbox{\tiny BFKL}}}\Rb\bar\rho^{\rm d.a.}_1\Lb Y;  \xi'_1\Rb \bar\rho^{\rm d.a.} _1\Lb Y;  \xi'_2\Rb\,=\,\Lb \frac{\bas}{\Delta_{\mbox{\tiny BFKL}}}\Rb G_{\pom}\Lb Y;  \xi'_1\Rb \,G_{\pom}\Lb Y;  \xi'_2\Rb \nn    \eea

where $\xi'_i$ is defined in \eq{SOLN1FM}. $\xi'_{ik}$ is the same as $\xi'_i$ where $r_i$ is replaced by $|\rv_i + \rv_k|$. For large values of $Y$ \eq{RHO2F}  has a very simple meaning shown in \fig{pom2}: it stems from the simple `fan' diagram after integration over $Y'$. Note, that in momentum representation the triple Pomeron vertex is equal to $\bas$.

~
   
     \begin{boldmath}
     \subsubsection{ $\bar{\rho}_3\Lb Y, r, r_1,b_1, r_2,b_2, r_3,b_2\Rb$}
      \end{boldmath}     
     From \eq{N301} we have the following equation for   $\bar{\rho}_3\Lb Y, k_T, r_1,b_1, r_2,b_2, r_3,b_2\Rb$ :
  \bea \label{EQN3}
  &&\int d^2\,r'  K\Lb \vec{r}',\vec{r} - \vec{r'}|\vec{r}\Rb
\Bigg\{ \Bigg(\,\bar \rho_3(Y,\rv', \bv -\h(\rv - \rv'); \rv_1, b_1, \rv_2, b_2,\rv_3,b_3)\,
+\,\,
 \bar \rho_3(Y,\rv - \rv', \bv -\h \rv'; \rv_1, b_1, \rv_2, b_2,\rv_3,b_3)\nn\\
 & -& \bar \rho_3(Y,\rv , \bv ; \rv_1, b_1, \rv_2, b_2) \Bigg)
 + \bar\rho_{1}(Y,\rv', \vec{b} ; \rv_1, \bv_1) \,\bar\rho_2(Y,\rv -\rv', \bv; \rv_2, \bv_2,\rv_3,b_3) \nn\\
 &\,+&\,\bar\rho_{2}(Y,\rv', \vec{b} ; \rv_1, \bv_1,\rv_3,b_3) \,\bar\rho_1(Y,\rv -\rv', \bv; \rv_2, \bv_2)\Bigg\}\,\, =\,\, \int\,\frac{d^2\,r'}{2\,\pi}\,
K\Lb \vec{r}',\vec{r}_1 - \vec{r'}|\vec{r}_1\Rb\,\\
&\times&\Big\{\bar{\rho}_3\Lb Y,\vec{k}_T;  \vec{r}',b_1,\vec{r}_2, b_2,\rv_3,b_3\Rb\,\,+\,\,\bar{\rho}_3\Lb Y,\vec{k}_T; \vec{r}_1 - \vec{r}',b_1,\rv_2, b_2, \rv_3,b_3\Rb\,-\,\bar{\rho}_3\Lb Y, \vec{k}_T, \rv_1,b_1, \rv_2,b_2, \rv_3,b_3\Rb\Big\}
\nn\\
&+&  \int\,\frac{d^2\,r'}{2\,\pi}\,
K\Lb \vec{r}',\vec{r}_2 - \vec{r'}|\vec{r}_2\Rb\,\nn\\
&\times&
\Big\{\bar{\rho}_3\Lb Y,\rv;  \vec{r}_1,b_1,\vec{r}', b_2, \rv_3,b_3\Rb\,\,+\,\,\bar{\rho}_3\Lb Y,\rv; \vec{r}_1,b_1,  \vec{r}_2 - \vec{r}',b_2,\rv_3,b_3  \Rb\,-\, \bar\rho_2(Y, \rv; \vec{r}_1, b_1,\vec{r}_2, b_2,\rv_3,b_3)\Big\}\nn\\
&+&\,\,\,\int\,\frac{d^2\,r'}{2\,\pi}\,
K\Lb \vec{r}',\vec{r}_3 - \vec{r'}|\vec{r}_3\Rb\,\nn\\
&\times&\Big\{\bar{\rho}_3\Lb Y,\rv;  \vec{r}_1,b_1,\vec{r}_2, b_2, \rv',b_3\Rb\,\,+\,\,\bar{\rho}_3\Lb Y,\rv; \vec{r}_1,b_1,  \rv_2,b_2,\vec{r}_3 - \vec{r}',b_3  \Rb\,-\, \bar\rho_2(Y, \rv; \vec{r}_1, b_1,\vec{r}_2, b_2,\rv_3,b_3)\Big\}
\nn\\
&+&\,\,
\,\, \,\Big\{  \bar\rho_2(Y, \rv;\vec{r}_1 +\vec{r}_3, b_1, \rv_2, b_2)\,\,+\,\,\bar\rho_2(Y, \rv;\vec{r}_1 , b_1,\vec{r}_2 +\vec{r}_3, b_2)\Big\}\nn \eea
\vspace{-0.2cm}
 Replacing $\int d^2\,r'  K\Lb \vec{r}',\vec{r} - \vec{r'}|\vec{r}\Rb\,\{\dots\} = \Delta_{\mbox{\tiny BFKL}}  \bar\rho_3(Y, \rv;   \rv_1, \bv_1, \rv_2, \bv_2, \rv_3, \bv_3)$  
  we obtain that
 \bea \label{RHO3EQ2}
  &&\bar\rho_3(Y, \rv;   \rv_1, \bv_1, \rv_2, \bv_2, \rv_3, \bv_3) \,\,=\,\,\h\Bigg\{ \frac{\bas}{\Delta_{\mbox{\tiny BFKL}}} \int d^2\,r'  K\Lb \vec{r}',\vec{r} - \vec{r'}|\vec{r}\Rb  \\
 &\times&  \Big\{ \bar\rho_{2}(Y, \rv', \vec{b} ; \rv_1, \bv_1, \rv_3, \bv_2 ) 
\,\,\bar\rho_{1}(Y,  \rv -\rv', \vec{b} ; \rv_2, \bv_2  ) \,\,+\,\,\bar\rho_{1}(Y, \rv', \vec{b} ; \vec{r}_1, \bv_1 ) 
\,\,\bar\rho_{2}(Y,  \rv - \rv', \vec{b} ; \rv_2, \bv_2, \rv_3, \bv_3  )\Big\}\,\,\nn\\
&& -\,\,\Big\{ \bar\rho_2(Y, \rv;\vec{r}_1 +\vec{r}_3, b_1, \rv_2, b_2)\,\,+\,\,\bar\rho_2(Y, \rv;\vec{r}_1 , b_1,\vec{r}_2 +\vec{r}_3, b_2)\Big\}\Bigg\}\nn\\
\eea
This equation has a simplified form in the momentum representation (see \eq{MMOM}):

\bea \label{RHO3F}
&&\bar\rho_3(Y, \kv;   \rv_1, \bv_1, \rv_2, \bv_2, \rv_3, \bv_3) \,\,=\,\,\h\Bigg( \frac{\bas}{\Delta_{\mbox{\tiny BFKL}}}\Bigg)\Bigg\{
   \bar\rho_{2}(Y, \kv, \vec{b} ; \rv_1, \bv_1, \rv_3, \bv_2 ) \bar\rho_{1}(Y,  \kv, \vec{b} ; \rv_2, \bv_2  )\,\,\\
   &&+\,\, \bar\rho_{1}(Y,  \kv, \vec{b} ; \rv_1, \bv_1  )\, \bar\rho_{2}(Y, \kv, \vec{b} ; \rv_2, \bv_2, \rv_3, \bv_2 ) \,\, -\,\,\Big( \bar\rho_2(Y, \kv;\vec{r}_1 +\vec{r}_3, b_1, \rv_2, b_2)\,\,+\,\,\bar\rho_2(Y, \kv;\vec{r}_1 , b_1,\vec{r}_2 +\vec{r}_3, b_2)\Big)\Bigg\}\nn
\eea

For the toy model in which $\bar \rho_n$ do not depend on the dipole sizes, \eq{RHO3F}
leads  to $\bar \rho_3\,\,=\,\,\bar \rho_1 \Lb \bar \rho_1\,-\,1\Rb^2$, if we assume that $\Delta_{\mbox{\tiny BFKL}}  \,=\,\bas$.    One can see that the calculations started to be cumbersome, but for our approach the most important conclusions is that the main contributions, which is proportional to $e^{ 3\,\Delta_{\mbox{\tiny BFKL}}\,Y}$, has  a very simple form:
 \bea\label{RHO3EQ3}
&& \bar\rho_3(Y, \kv;   \rv_1, \bv_1, \rv_2, \bv_2, \rv_3, \bv_3)\,\,=\\
&&\,\, \Bigg( \frac{\bas}{\Delta_{\mbox{\tiny BFKL}}}\Bigg)^2 \bar\rho_{1}(Y,  \kv, \vec{b} ; \rv_1, \bv_1  )\,\bar\rho_{1}(Y, \kv, \vec{b} ; \rv_2, \bv_2  )\,\bar\rho_{1}(Y,  \rv, \vec{b} ; \rv_3, \bv_3  )\,\,+\,\,{\cal O}\Lb e^{ 2\,\Delta_{\mbox{\tiny BFKL}}\,Y}\Rb\Bigg\}\nn\\
 &&\,\,\Bigg( \frac{\bas}{\Delta_{\mbox{\tiny BFKL}}}\Bigg)^2 \bar\rho^{\rm d.a}_1\Lb Y;  \xi'_1\Rb\, \bar\rho^{\rm d.a.}_1\Lb Y;  \xi'_2\Rb\,\bar\rho^{\rm d.a.}_1\Lb Y;  \xi'_3\Rb\,\,=\,\,\Lb \frac{\bas}{\Delta_{\mbox{\tiny BFKL}}}\Rb^2 G_{\pom}\Lb Y;  \xi'_1\Rb \,G_{\pom}\Lb Y;  \xi'_2\Rb\,G_{\pom}\Lb Y;  \xi'_3\Rb 
  \nn
 \eea
We see again that at large $Y$  $\bar \rho_3$ is described by the `fan' diagram of the same type as in \fig{pom2}, but with three Pomerons,  in which two integrations of the positions of two triple Pomeron vertices lead to factor $(1/\Delta_{\mbox{\tiny BFKL}})^2$ while $\bas^2$ appears due to the value of the triple Pomeron vertex is equal to $\bas$ and we have two vertices in these diagrams.

   ~

     \begin{boldmath}
     \subsubsection{ $\bar{\rho}_n\Lb Y, \rv,\{\rv_i,b_i\}\Rb$ at high energies}
      \end{boldmath}

     Having  solutions for $\bar \rho_2$ and $\bar \rho_3$, one can see that the  leading term of the solution to \eq{N301} , which behaves as $ \exp\Lb n \Delta_{\mbox{\tiny BFKL}}\,Y\Rb $ at high energies, has the form:
  \beq \label{RHONLT}
  \bar{\rho}_n(Y,\{ \xi'_i\} )\,\,=\,\, \Bigg( \frac{\bas}{\Delta_{\mbox{\tiny BFKL}}}\Bigg)^{n-1}  
       \underbrace{\prod^n_{i=1}  \bar\rho^{\rm d.a.} _1\Lb Y;  \xi'_i\Rb}_{\sim \exp\Lb n \Delta_{\mbox{\tiny BFKL}}\,Y\Rb}   
\eeq

     We can check by the direct substitution in \eq{N301} that  $\bar \rho_n $ is equal to
   \beq \label{SOLNN} 
   \bar{\rho}_n(Y,\{ \xi_i\} )\,\,=\,\,  \Bigg( \frac{\bas}{\Delta_{\mbox{\tiny BFKL}}}\Bigg)^{n-1}  \Bigg\{
      \underbrace{\prod^n_{i=1}  \bar\rho^{\rm d.a.} _1\Lb Y;  \xi_i\Rb}_{\sim \exp\Lb n \Delta_{\mbox{\tiny BFKL}}\Rb} \,\,-\,\,\frac{1}{n-1}\underbrace{\sum_{i=1}^{n-1}  \bar\rho^{\rm d.a.}_{n-1}(\{\xi_j ,\xi_{i,n}\})}_{\sim \exp\Lb( n-1) \Delta_{\mbox{\tiny BFKL}}\Rb}\Bigg\}
      \eeq
    Accuracy of this solution is of the order of $n\,\Lb \xi^2_i/(4\,D\,Y)\Rb^2$ and we will show below that the typical values of $n$ does not increase with $Y$.

   ~

     \begin{boldmath}
     \section{ Scattering amplitude}
      \end{boldmath}

    
     \begin{boldmath}
     \subsection{The BFKL Pomeron exchange}
      \end{boldmath}

    From \eq{MPSI} the contribution of the single  BFKL Pomeron exchange to the scattering amplitude is equal to the following expression\cite{MUDI}:
    \beq \label{BFKL1}
    A^{\mbox{\tiny BFKL}}\Lb Y, \kv, \kv'; \vec{b}\Rb\,\,=\,\,\int \frac{d^2\,r_1}{r^4_2}  \frac{d^2\,r'_1}{r'^4_1} \,d^2 b'_1\,\bar\rho\Lb  Y  - Y_0, \kv;  \rv_1, \vec{b} - \vec{b}'_1\Rb \,\,\sigma^{\rm BA}\Lb \rv_1, \rv'_1 \Rb  \,\, \bar\rho\Lb   Y_0, \kv';  \rv'_1,  \vec{b}'_1\Rb 
    \eeq
    where $\kv'$  appears as  momentum variable corresponding the $R$-dependece of the scattering amplitude.

   \eq{BFKL1} has been estimated (see Refs.\cite{MUDI,Salam}) , however, for completeness of presentation,  we briefly outline  here the main points of these estimates. Plugging \eq{SOLN1} and \eq{BAAMPL} into \eq{BFKL1} we can first integrate over $r_1$ and $r'_1$, obtaining the contribution $\propto\,\,l^{2+ 2 i \nu\,+\,2\,i\,\nu'}$, with $\gamma_r=\h + i\,\nu$ and $\gamma'_{r'} = \h + i \nu'$.  Integration over $l$ leads to the pole $1/(\nu\,+\nu')$, which contribution leads to the independence of the amplitude with the value of $Y_0$. The integral over $d ^2 b'_1$  has the form
   
   \beq \label{BFKL2}
 \int d^2 b'_1 \frac{1}{\Lb \Lb \vec{b} \,-\,\vec{b}'_1\Rb^2\,b'^2_1\Rb^{1 + i \,\nu} }\,\,=\,\,\frac{2}{i\nu} \frac{1}{b^{2 (1 + i\nu)}} \eeq
   In \eq{BFKL2} the main contributions stem from two kinematic regions: $b'_1 \ll b$ and $| \vec{b} - \vec{b}'_1| \,\ll\,b$. Finally we obtain the result of Ref.\cite{MUDI}:
      \bea \label{BFKL3}
   && A^{\mbox{\tiny BFKL}}\Lb Y, \kv, \kv'; \vec{b}\Rb\,\, =\\\
   &&\frac{ 8 \bas^2\,\pi}{\Lb D\,Y\Rb^{3/2}} \,\frac{4}{k_T\,k'_T\,b^2} \,\ln\Lb \frac{k_T\,k'_T\,b^2}{4} \Rb  
   \,\exp\Lb  \Delta_{\mbox{\tiny{BFKL}}}\,Y-\frac{ \ln^2\Lb\frac{16}{b^4\,k^2_T\,k'^2_T}\Rb}{ 4\,D\,Y}\Rb\,\,=\,\,\frac{ 8 \,\pi \bas^2
   }{\Lb D\,Y\Rb^{3/2}} \,\xi\,e^{-\xi}\,\exp\Lb  \Delta_{\mbox{\tiny{BFKL}}}\,Y-\frac{ \xi^2}{4\,D\,Y}\Rb\nn\eea
    where $\xi = \ln\Lb \frac{k_T\,k'_Tb^2}{4} \Rb $.
    
    ~

     \begin{boldmath}
     \subsection{Scattering amplitude from the main formula in the momentum representation}
      \end{boldmath}

         Plugging \eq{BFKL3} into our master equation (see \eq{MPSI}) one can see that the scattering amplitude takes the following form:  
   \bea \label{MPSI1}
  &&A\Lb Y, \kv,\kv' ;  \vec{b}\Rb\,=\,\sum^\infty_{n=1}\,\Lb -1\Rb^{n+1}\,n!\,  \Bigg(\frac{{\bas^2}}{\Delta^2_{\mbox{\tiny BFKL}}} \Bigg)^{n -1} \Lb   A^{\mbox{\tiny BFKL}}\Lb Y, k_T, k'_T; \vec{b}\Rb\Rb^n\nn\\ 
 && \,\,=\,\,\Bigg(\frac{\Delta^2_{\mbox{\tiny BFKL}}}{\bas^2} \Bigg)\Bigg\{1\,\,-\,\,  e^{\frac{1}{\kappa} }\,\Gamma\Lb 0,\frac{1}{\kappa}\Rb/\kappa\Bigg\} \,\,\rightarrow\,\,
 \left\{\begin{array}{l}\,\,\,\Big(\frac{\Delta^2_{\mbox{\tiny BFKL}}}{\bas^2} \Big)\,\kappa = A^{\mbox{\tiny BFKL}}\Lb Y, r, R; \vec{b}\Rb \,\,\,\,\,\,\,\,\,\mbox{for}\,\,\,\kappa\,\ll\,1;\\ \\
\,\,\Big(\frac{\Delta^2_{\mbox{\tiny BFKL}}}{\bas^2} \Big)\Bigg(1\,+\frac{-\ln \kappa\,\,+C_E}{\kappa}\Bigg)\,\,\,\mbox{for}\,\,\,\kappa\,\gg\,1;\\  \end{array}
\right.
  \eea         
   where $\kappa =   \Big(\frac{\bas^2}{\Delta^2_{\mbox{\tiny BFKL}}} \Big) A^{\mbox{\tiny BFKL}}\Lb Y, \kv, \kv'; \vec{b}\Rb$. One can see that at low energies (small values of $Y$) the scattering amplitude reproduces the exchange of one BFKL Pomeron. At high energies ( at $Y\,\,\gg\,\,1$) the amplitude  approaches the constant value $\Big(\frac{\Delta_{\mbox{\tiny BFKL}}}{\bas} \Big)$. Since this amplitude is in the momentum representation, the unitarity limit for it is $\h \ln k^2_T\h \ln k'^2_T  $, but not the unity.

    The scattering amplitude of    \eq{MPSI1}   does not 
    generate a correct behaviour of the cross section, which increases as a power of the energy, resulting from the power-like bahaviour of the scatterring amplitude at large impact parameters. From \eq{MPSI1} one can see that the corrections at large values of $Y$ show the increase with the growth of $b$, demonstrating that in the scattering amplitude we have even more severe problems with the Froissart theorem\cite{FROI} than for the BK evolution\cite{KW1,KW2,KW3,FIIM}.

    In all our estimates we assume that $\sum^n_1 \frac{\xi'_i}{2\,D,Y} \,\,\ll\,\,1$.  Therefore, we need to estimate the typical values of $n$ in this sum.
  From  \eq{MPSI1} we can find the average value of  $n$:
    \beq \label{MPSI2}
    \bar n\,=\frac{d\, \ln A\Lb Y, r,R ;  \vec{b}\Rb}{d \ln A^{\rm BFKL} }  =\,\,\Bigg(\frac{\mu e^{1/\kappa } \Gamma \left(0,\frac{1}{\kappa }\right)}{\kappa ^2}-\frac{\mu}{\kappa }+\frac{\mu e^{1/\kappa } \Gamma \left(0,\frac{1}{\kappa }\right)}{\kappa }\Bigg)\Bigg{/}\Bigg(1-\frac{\mu e^{1/\kappa } \Gamma \left(0,\frac{1}{\kappa }\right)}{\kappa }\Bigg)    \eeq

      where $\kappa =   \Big(\frac{\bas^2}{\Delta^2_{\mbox{\tiny BFKL}}} \Big) A^{\mbox{\tiny BFKL}}\Lb Y, r, R; \vec{b}\Rb$ and $\mu = \Delta^2_{\mbox{\tiny BFKL}}/{\bas^2}$. One can see that  $\bar n$ decreases at large values of $Y$. Therefore, our assumption  $\sum^n_1 \frac{\xi'_i}{2\,D,Y} \,\,\ll\,\,1$ looks plausible.  
      
      One can also see that at fixed $b$,  the scattering amplitude approaches  the constant value of $\Delta_{\mbox{\tiny BFKL}}/{\bas}$       as follows
  \beq \label{MPSI3}
   A\Lb Y,\kv,\kv' ;  \vec{b}\Rb\,\,=\,\,1\,\,-\,\,{\rm Const}\ln \Lb A^{\mbox{\tiny BFKL}}\Lb Y, r, R; \vec{b}\Rb\Rb/ A^{\mbox{\tiny BFKL}}\Lb Y, \kv, \kv'; \vec{b}\Rb\,\,=\,\,1\,-\,{\cal C}\Lb \xi',Y\Rb \Lb e^{\xi' }\,Q^2_s(Y)\Rb^{- \bar \gamma}
 \eeq
 where  the saturation momentum $Q^2_s(Y) \,=\,\exp\Lb 2\,D\,Y \Lb 1 + \frac{\Delta_{\mbox{\tiny BFKL}} }{D}\Rb\Rb$,$\bar \gamma \,=\,\h + \frac{ \Delta_{\mbox{\tiny BFKL}}}{D}$ and function ${\cal C}$ is a smooth function of $\xi$ and $Y$.
 
    It is instructive to note that at first sight such approach is in contradiction both with BK non-linear evolution equation\cite{LETU} and with estimates for the scattering dipole-dipole amplitude in Ref. \cite{IAMU1}. However, we will show in the next section that it is not the case, considering the solution to BK non-linear equation in the diffusion approximation for the BFKL kernel.
    
    ~

    ~~
         ~
     \begin{boldmath}
     \subsection{Solution to  BK equation with the diffusion kernel at high energies}
      \end{boldmath}


    The surprising result is that the amplitude in the momentum representation turns out to be constant at high energies and it does not show the geometric scaling behaviour. In this section we show that both these features are the artifact of the  simplified  BFKL kernel that we have used.  As have been discussed we used the diffusion approximation  of \eq{CHI} (at $\gamma \to \h$) for the BFKL kernel. 
    The BK equation in the momentum representation ($k_T$) takes the following form:
\beq \label{BKMR}
\frac{\partial \widetilde{N}\Lb k_\perp, b; Y\Rb}{\partial Y}\,\,=\,\,\bas \Bigg\{ \chi\Lb -\,\frac{\partial}{\partial \tilde{\xi}}\Rb
\widetilde{N}\Lb k_\perp, b; Y\Rb\,\,\,-\,\, \widetilde{N}^2\Lb k_\perp, b; Y\Rb\Bigg\}
\eeq
  with $  \xt = \ln k^2_T$ and 
  \beq \label{HA1}
 \bas \,\chi\Lb -\,\frac{\partial}{\partial \tilde{\xi}}\Rb\,\,=\,\,\Delta_{\mbox{\tiny BFKL}} \,\,+\,\,D\,\frac{
  \partial^2}{\partial \xt^2} 
  \eeq
  (see \eq{CHI} at $\gamma \to \h$).
  
  The asymptotic solution to \eq{BKMR} has a simple form:  $\widetilde{N}^{\rm asymp}\Lb k_\perp\Rb  = \frac{ \Delta_{\mbox{\tiny BFKL}}}{\bas}$.  Plugging in \eq{BKMR} $ \widetilde{N}\Lb k_\perp, b; Y\Rb \,\,=\,\,
\widetilde{N}^{\rm asymp}\Lb k_\perp\Rb \,\,-\,\,\Delta\widetilde{N}\Lb k_\perp, b; Y\Rb $ and considering  
$\Delta\widetilde{N}\Lb k_\perp, b; Y\Rb \,\,\ll\,\,1$  we obtain the following linear equation for $\Delta\widetilde{N}\Lb k_\perp, b; Y\Rb$:
\beq \label{BKMR1}
\frac{\partial \Delta\widetilde{N}\Lb k_\perp, b; Y\Rb}{\partial Y}\,\,=\,\,- \Delta_{\mbox{\tiny BFKL}} \Delta\widetilde{N}\Lb k_\perp, b; Y\Rb\,\,+\,\,D\,\frac{
  \partial^2}{\partial \xt^2}\, \Delta\widetilde{N}\Lb k_\perp, b; Y\Rb
\eeq
 \eq{BKMR1} has the same form as the linear BFKL equation but with  the negative intercept. Hence, the solution to \eq{BKMR1}  can be obtain from \eq{SOLN1FM} :

 \beq \label{BKMR2}
\Delta\widetilde{N}\Lb k_\perp, b; Y\Rb\,\,=\,\,\frac{2}{\Lb D\,Y\Rb^{3/2}}\,\xi'\,e^{-\Delta_{\mbox{\tiny{BFKL}}}\,Y-\frac{\xi'^2}{4\,D\,Y}}\,e^{\h \xi'}
\eeq  
 
 Therefore, one can see that (i) the solution at high energies does not show the geometric scaling behaviour which was predicted in Ref.\cite{LETU}; and (ii) at large $Y$  it decreases as $ \exp\Lb - {\rm Const} Y\Rb$ instead of $\exp\Lb - {\rm Const} Y^2\Rb$\cite{LETU}. Note, that we use the same procedure  to \eq{BKMR} as was developed in Ref.\cite{LETU} to the general kernel of the BFKL equation. 
 On the other hand, applying the approach of Ref.\cite{IAMU1} to the scattering amplitude  taking into account \eq{BKMR2} we obtain \eq{MPSI1} for $\kappa \gg 1$.
 
 Concluding, we state that the scattering amplitude of \eq{MPSI1} gives more microscopic insight in the structure of the scattering amplitude and reproduces both approaches of Refs.\cite{IAMU,LETU}. It worth mentioning that \eq{BKMR}, being F-KPP equations\cite{F,KPP}, shows the geometrical scaling behaviour in pre-asymptotic region in the vicinity of the saturation scale. \eq{BKMR2} also indicates that  our assumption:  $\xi/Y\,\,\ll\,1$ is not substantial for the main features of the amplitude behaviour.
 
   ~

   ~
    
     \begin{boldmath}
     \subsection{Dipole-dipole scattering  amplitude at high energies }
      \end{boldmath}


    The dipole-dipole  scattering amplitude takes the following form:
    \beq \label{SA1}
    A\Lb Y, r, R ;  \vec{b}\Rb  \,\,=\,\,r^2\,R^2\,\int k_T \,dk_T J_0\Lb k_T\,r\Rb\, \int k'_T \,dk'_T J_0\Lb k'_T\,R \Rb  \,\,A\Lb Y, \kv,\kv' ;  \vec{b}\Rb
    \eeq
    We use  the Mellin transform for $J_0(k_T\, r)$ (see formula 6.8.1 of Ref.\cite{BATEMAN} ) :
     \beq \label{SA2} 
     J_0(k_T\, r)\,\,=\,\,\int^{\epsilon + i \infty}_{\epsilon - i \infty} \frac{d \gamma}{2\,\pi\,i} (k_T\,r)^{- \gamma} \frac{2 ^{\gamma - 1} \Gamma\Lb \h \gamma\Rb}{\Gamma\Lb1 - \h \gamma\Rb}
     \eeq     
    Note, that in \eq{SA2} we take $\frac{3}{2}  > \epsilon > 1$. 
     
 Introducing a new variable $k_T\,k'_t \,=\zeta$ we can rewrite \eq{SA1} in the form:
   \bea \label{SA3}
     &&A\Lb Y, r, R ;  \vec{b}\Rb  \,\,=\\
     &&\,\,r^2\,R^2\,\int \zeta d \zeta \frac{d k_T}{k_T}\,\int^{\epsilon + i \infty}_{\epsilon - i \infty} \frac{d \gamma}{2\,\pi\,i} (k_T\,r)^{- \gamma} \frac{2^{\gamma - 1} \Gamma\Lb \h \gamma\Rb}{\Gamma\Lb1 - \h \gamma\Rb}\int^{\epsilon + i \infty}_{\epsilon - i \infty} \frac{d \gamma}{2\,\pi\,i} \Lb\frac{\zeta}{k_T}\,R\Rb^{- \gamma'} \frac{2^{\gamma' - 1} \Gamma\Lb \h \gamma'\Rb}{\Gamma\Lb1 - \h \gamma'\Rb}
  \,\,A\Lb Y, \kv,\kv' ;  \vec{b}\Rb\nn 
  \eea   
    Noting that  $ A\Lb Y, \kv,\kv' ;  \vec{b}\Rb  $ depends only on $\zeta$, we can integrate \eq{SA3} over $k_T$ and $\gamma'$ ,  which results in:
    
    \beq \label{SA4}
     A\Lb Y, r, R ;  \vec{b}\Rb  \,\,=\,\,r^2\,R^2\,\int \zeta d \zeta \,\int^{\epsilon + i \infty}_{\epsilon - i \infty} \frac{d \gamma}{2\,\pi\,i} \Lb\zeta\,R\,r\Rb^{- \gamma} \frac{4^{\gamma - 1} \Gamma^2\Lb \h \gamma\Rb}{\Gamma^2\Lb1 - \h \gamma\Rb}
  \,\,A\Lb Y, \kv,\kv' ;  \vec{b}\Rb  
  \eeq         
    
    To take integral over $\zeta$  we simplify the expression for 
   $ A^{\mbox{\tiny BFKL}}\Lb Y, \kv, \kv'; \vec{b}\Rb  $ of \eq{BFKL3}  considering     
        $\gamma_{SP}=\h$ and  reducing it to the following expression:
        
        \bea \label{SA5}
         A^{\mbox{\tiny BFKL}}\Lb Y, \kv, \kv'; \vec{b}\Rb\,\,&=&\,\,\frac{ 8 \,\pi}{\Lb D\,Y\Rb^{3/2}} \,\frac{4}{k_T\,k'_T\,b^2} \,\ln\Lb \frac{k_T\,k'_T\,b^2}{4} \Rb  
   \,\exp\Lb  \Delta_{\mbox{\tiny{BFKL}}}\,Y-\frac{ \ln^2\Lb\frac{16}{b^4\,k^2_T\,k'^2_T}\Rb}{ 4\,D\,Y}\Rb\nn\\
   &\,\,=\,\,&\frac{1}{k_T\,k'_T} N\Lb k_T \to 2/r; k'_T \to 2/R\Rb\,\,=\,\,\frac{N}{\zeta}
   \eea        
   where we use \eq{SOLN1FM}. The new variable $\zeta = k_T\,k'_T$.
   
   Using  the integral representation for the incomplete gamma function (see  formula {\bf 8.353(3)} in Ref.\cite{RY}) and \eq{SA5} we obtain:

\bea \label{SA6}
  A\Lb Y, r, R ;  \vec{b}\Rb  \,\,&=&\,\,\mu\,r^2\,R^2\,\int \zeta d \zeta \,\int^{\epsilon + i \infty}_{\epsilon - i \infty} \frac{d \gamma}{2\,\pi\,i} (\zeta\,R\,r)^{- \gamma} \frac{4^{\gamma - 1} \Gamma^2\Lb \h \gamma\Rb}{\Gamma^2\Lb1 - \h \gamma\Rb}\,\int^\infty_0 d \tau e^{- \tau}  \frac{\frac{\tau\,N}{\mu\,\zeta} }{1 +\frac{\tau\,N}{\mu\,\zeta}}\nn\\
  &=&\mu\, r^2\,R^2\,\int \zeta d \zeta \,\int^{\epsilon + i \infty}_{\epsilon - i \infty} \frac{d \gamma}{2\,\pi\,i} (R\,r)^{- \gamma} \frac{4^{\gamma - 1} \Gamma^2\Lb \h \gamma\Rb}{\Gamma^2\Lb1 - \h \gamma\Rb}\,\int^\infty_0 d \tau e^{- \tau}\Bigg\{ - \frac{\pi}{\sin\Lb\pi \,\gamma\Rb}\Lb \tau\,N/\mu\Rb^{(2 - \gamma)}\Bigg\} \nn\\
  &=& \mu\int^{\epsilon + i \infty}_{\epsilon - i \infty} \frac{d \gamma}{2\,\pi\,i}  \frac{4^{\gamma - 1} \Gamma^2\Lb \h \gamma\Rb}{\Gamma^2\Lb1 - \h \gamma\Rb} \Bigg( - \frac{\pi}{\sin\Lb \pi \,\gamma\Rb} \Bigg) \Lb r\,R\, N/\mu\Rb^{2 - \gamma}
   \eea

    Closing contour of integration over $\gamma$ on negative $\gamma$ we obtain the scattering amplitude as sum of $\Lb r \,R\, N  =  A^{ \mbox{\tiny BFKL}} \Lb Y, \xi_r\Rb\Rb^n$ , where $\xi_r= \ln\Lb \frac{b^2}{r\,R} \Rb$.   If we close the contour  on the singularities for positive $\gamma > 1$, we have the asymptotic series , which determines the behaviour of the scattering amplitude at high energies (at large values of $Y$). One can see that the integrant has no singularities at $\gamma = 2$ and the first pole appears at $\gamma = 3$, which leads to 
    
    \beq \label{SA7}
  A\Lb Y, r, R ;  \vec{b}\Rb\,\,=\,\,\mu^2\frac{\ln\Lb  A^{ \mbox{\tiny BFKL}} \Lb Y, \xi_r\Rb/\mu\Rb}{  A^{ \mbox{\tiny BFKL}} \Lb Y, \xi_r\Rb}  
  \eeq
  where $\mu\,  = \,\Delta_{\mbox{\tiny BFKL}}/{\bas}$.

    Therefore, we found that the scattering amplitude  decreases at large values of $Y$ (at high energies).

    ~  
      
      ~

     \begin{boldmath}
     \section{Conclusions }
      \end{boldmath}

      
      In this paper we have three results. First, we derived the recurrence relations for dipole  densities ($\rho_n$) in QCD for the BFKL parton cascade. These relations allow us to find the dipole densities from the solution to the BFKL equation for $\rho_1$. Note, that \eq{RHONNEQ}  for the energy evolution of the parton densities is also new. Second, using the diffusion approximation for the BFKL kernel, we resolve these recurrence relations and found  the leading terms in $\rho_n \propto e^{n\, \Delta_{\mbox{\tiny BFKL}} \,Y}$. It is worth mentioning that these relations are suited for the numerical estimate  of the  dipole densities opening a new way for the numerical simulation of the scattering amplitudes using \eq{MPSI}.
      
Third, for the first time we sum analytically  the large Pomeron loops in QCD  using these solutions. As a result of this summation we obtain the  dipole-dipole scattering amplitude. Surprisingly,  it turns out that this amplitude decreases at large values of $Y$.  We believe that such a behaviour of the scattering amplitude follows from the simplified kernel  of the BFKL equation in the diffusion  approximation,  as has been demonstrated in section IV-C. The physics origin of such behaviour  is that the diffusion BFKL kernel does not lead to the saturation both in BK equation and in dipole-dipole amplitude.   In other words, the first attempt to sum  analytically  BFKL Pomeron loops in QCD leads  to  the scattering amplitude that satisfies both $t$ and $s$ channel unitarity without saturation. Hence, the sum of Pomerom loops gives the typical contribution to S-matrix  at high energies which turns out to be larger than the rare fluctuations discussed in Ref\cite{IAMU1}  and which  will  lead to the main contribution to the scattering amplitude for more realistic approximation for the BFKL kernel. The fact that the diffusion approximation to the BFKL kernel is so deficient,  turns out  to  be a great surprise to us, especially because this approximation, which leads to F-KPP equation, has been widely used to describe the deep inelastic scattering (see Ref.\cite{KOLEB} for review). On the other hand,  such a result is not new for the Pomeron calculus (see Ref.\cite{BRAUN1}  and reference therein). It should be emphasized  that shortcomings of the diffusion approximation force us to look at numerical estimates of Refs.\cite{MUSA,Salam} with a grain of salt, since  the diffusion approximation was used in these papers. We have to believe that these estimates have been made in the pre-asymptotic region where the diffusion approximation generates the geometric scaling behaviour of the scattering amplitude.

Certainly, summing the Pomeron loops for more realistic approximation for the BFKL kernel, that leads to the saturation, will be our first problem to solve in the future. We wish also to note that even in present form the sum of Pomeron loops can be useful in discussion of the multiplicity distribution of the produced gluons.

We believe that our reader can take the following results  from this paper. First, it is the new evolution equation  for dipole densities (see \eq{RHONNEQ} and  the recurrence relation between tthem (see \eq{N301}). They   are derived for the general BFKL kernel. The recurrence relations are well suited for the numerical estimates of the scattering amplitudes. Second, it is   the solution of \eq{SOLNN}, which have been found for the  BFKL kernel in diffusion approximation. However, one can  use these solutions  only in the vicinity of the saturation scale, where they  reproduce the geometric scaling behaviour\cite{F,KPP}. The third is the unexpected result that the diffusion approximation cannot describe the high energy asymptotic behaviour  both for BK equation  and for dipole-dipole  scattering. The failure of the diffusion approximation is surprising and instructive since  most of experts bear in mind the diffusion approximation discussing the BFKL Pomeron contribution.

      ~
      
       {\bf Acknowledgements} 
     
   We thank our colleagues at Tel Aviv university and UTFSM for
 encouraging discussions. Special thanks go A. Kovner and M. Lublinsky for stimulating and encouraging discussions on the subject of this paper. This research    was supported  by 
 ANID PIA/APOYO AFB180002 (Chile),  Fondecyt   grant \#1180118 (Chile) and the Tel Aviv university encouragement grant \#5731.

\end{document}